\documentclass{emulateapj}
\newcommand{\teff}{$T_{\rm{eff}}$ }

\slugcomment{Published in The Astronomical Journal (AJ, 143, 103).
Available At: \url{http://stacks.iop.org/1538-3881/143/103}}

\shorttitle{Multi-Survey New White Dwarfs}
\shortauthors{Sayres et al.}

\begin{document}

\title{A Multi-Survey Approach to White Dwarf Discovery}


\author{Conor Sayres\altaffilmark{1,4,5}}\email{csayres@u.washington.edu}
\author{John P. Subasavage\altaffilmark{2,4}}
\author{P. Bergeron\altaffilmark{3}}
\author{P. Dufour\altaffilmark{3}}
\author{James R. A. Davenport\altaffilmark{1}} 
\author{Yusra AlSayyad\altaffilmark{1}} 
\author{Benjamin M. Tofflemire\altaffilmark{1}} 

%\author{Conor Sayres\altaffilmark{1,2}} 
%\affil{Department of Astronomy, University of Washington, Box 351580, Seattle, WA  98195}
%\email{csayres@u.washington.edu}
%
%\author{John P. Subasavage}
%\affil{Cerro Tololo Inter-American Observatory, Casilla 603, La Serena, Chile}
%
%\author{P. Bergeron}
%\affil{D\'{e}partement de Physique, Universit\'{e} de Montr\'{e}al, C.P.6128, Succursale Centre-Ville, Montr\'{e}al, QC H3C 3J7, Canada}
%
%\author{P. Dufour}
%\affil{D\'{e}partement de Physique, Universit\'{e} de Montr\'{e}al, C.P.6128, Succursale Centre-Ville, Montr\'{e}al, QC H3C 3J7, Canada}
%
%\author{James R. A. Davenport} 
%\affil{Department of Astronomy, University of Washington, Box 351580, Seattle, WA  98195}
%
%\author{Yusra AlSayyad} 
%\affil{Department of Astronomy, University of Washington, Box 351580, Seattle, WA  98195}
%
%\and
%
%\author{Benjamin M. Tofflemire} 
%\affil{Department of Astronomy, University of Washington, Box 351580, Seattle, WA  98195}
%

\altaffiltext{1}{Department of Astronomy, University of Washington, Box 351580, Seattle, WA  98195}
\altaffiltext{2}{US Naval Observatory, 10391 West Naval Observatory Road, Flagstaff, AZ 86001-8521}
\altaffiltext{3}{D\'{e}partement de Physique, Universit\'{e} de Montr\'{e}al, C.P.6128, Succursale Centre-Ville, Montr\'{e}al, QC H3C 3J7, Canada}
\altaffiltext{4}{Visiting Astronomer, Cerro Tololo Inter-American Observatory.
CTIO is operated by AURA, Inc.\ under contract to the National Science
Foundation.}
\altaffiltext{5}{CTIO REU 2010.}


\begin{abstract}

By selecting astrometric and photometric data from the Sloan Digital
Sky Survey (SDSS), the L{\'e}pine \& Shara Proper Motion North Catalog
(LSPM-North), the Two Micron All Sky Survey (2MASS), and the USNO-B1.0
catalog, we use a succession of methods to isolate white dwarf
candidates for follow-up spectroscopy. Our methods include: reduced
proper motion diagram cuts, color cuts, and atmospheric model
adherence.  We present spectroscopy of 26 white dwarfs obtained from
the CTIO 4m and APO 3.5m telescopes.  Additionally, we confirm 28
white dwarfs with spectra available in the SDSS DR7 database but
unpublished elsewhere, presenting a total of 54 WDs.  We label one of
these as a recovered WD while the remaining 53 are new discoveries.
We determine physical parameters and estimate distances based on
atmospheric model analyses.  Three new white dwarfs are modeled to lie
within 25 pc.  Two additional white dwarfs are confirmed to be
metal-polluted (DAZ).  Follow-up time series photometry confirms
another object to be a pulsating ZZ Ceti white dwarf.

\end{abstract}


\keywords{Catalogs - Proper motions  - Stars: variables: general - Surveys - white dwarfs}

\section{Introduction}

White dwarfs (WDs) are of interest in a variety of subfields in
astrophysics because of their unique ability to act as cosmic
chronometers.  Theoretical cooling models provide a means for dating
WDs from photometry alone, and this process has provided constraints
on the age of the Galactic disk using cool WD samples
\citep[e.g.,][]{leggett}.  Because the coolest (and hence oldest) WDs
are less luminous, the nearby representatives provide the best
opportunities for accurate characterization.

The present understanding of the local WD sample is somewhat
uncertain.  \citet{holberg_nearby} estimate that the WD sample is
$\sim$80\% complete to 20 pc.  Including only WDs with accurate
trigonometric parallaxes, \citet{subasavage} conclude that the sample
is only $\sim$47\% complete to 25 pc.  The vast majority of the
incompleteness arises from the coolest WDs that remain to be
discovered.  Additional discoveries of nearby WDs will strengthen the
completeness statistics and provide valuable model parameter
constraints through trigonometric parallax measurements.

The Sloan Digital Sky Survey (SDSS; \citealt{york}) has been
exceptional at identifying WDs.  The lastest release as of the writing
of this manuscript, Data Release 8 (DR8; \citealt{dr8}), reports sky
coverage of over 14,500 deg$^2$.  Imaging data are collected in the
$ugriz$ bands with a 50\% completeness limit at $r$ = 22.5 for point
sources, and spectra have been observed for over half of a million
stars.  Prior data releases from SDSS have led to a proliferation of
WD discoveries.  \cite{eisenstein} (hereafter SDSS-E06) roughly
doubled the number of previous, spectroscopically identified WDs using
spectra from the DR4 database \citep{dr4}, though most were hotter
than $\sim$7000 K because of biases in the SDSS spectroscopic target
selection process.  The observational efforts of
\cite{kilic2006,kilic2010} led to spectroscopic confirmation of more
than 100 cool WDs, where target selection was based on photometry and
proper motions contained in the SDSS DR7 database \citep{dr7}.

In this work, we utilize data from a suite of surveys and catalogs,
including SDSS, the Two Micron All Sky Survey \citep[2MASS;
][]{2mass}, USNO-B1.0 \citep{monet}, and the proper motion survey of
\cite{lspm} to identify new WD candidates.  From these data, we
prioritized targets for follow-up spectroscopy, with emphasis on cool,
nearby WDs as well as hotter, and potentially variable (i.e., ZZ Ceti)
WDs.  We present spectra and physical parameters, derived from
spectral energy distribution (SED) and model atmosphere analyses, for
25 newly discovered WDs and one recovered WD (see \ref{subsec:notes}).
In addition, we present physical parameters for 28 new WDs that have
been spectroscopically confirmed by SDSS DR7 spectra but are
unpublished elsewhere.

\section{Target Selection}
\label{sec:TargSel}

\subsection{Methodology}
\label{subsec:Method}

Target selection was based on a combination of constraints applied to
optical and near-IR photometry, proper motion, and model-adherence.
Two independent target selection passes were employed due to high
contamination rates experienced with the initial approach. We first
queried the SDSS DR7 database for objects with the following
stipulations: proper motion $> 100$ mas yr$^{-1}$, $g$ magnitude $<
19.5$, and declination $< +30^\circ$.  The declination constraint was
used to ensure all targets could be observed from Cerro Tololo
Inter-American Observatory (CTIO), where the first round of
spectroscopic observations was taken.  \cite{munn} have joined the
USNO-B1.0 Catalog with SDSS astrometry (available in the {\em
  propermotions} table of the SDSS database), which allowed us to
query for high proper motion objects directly from SDSS.  In a second
pass, we began with proper motion objects from the catalog of
\cite{lspm} containing objects with proper motions
$\geq$150 mas yr$^{-1}$.  For these objects, we required a match in
SDSS and employed the $g$ magnitude $< 19.5$ limit with no declination
constraint.  In both approaches, candidate objects were required to
have near-IR photometry from 2MASS.  Cross-matching between catalogs
was achieved using the table matching routines of
TOPCAT\footnote{\url{http://www.star.bristol.ac.uk/~mbt/topcat/}}
\citep{2005ASPC..347...29T}.

The technique of using reduced proper motion (RPM) to identify WDs has
been used for decades \citep[e.g., ][]{1972ApJ...177..245J}.
Conceptually, RPM is used as a proxy for absolute magnitude.  The two
quantities are connected by the inclusion of tangential velocity.
Thus, proper motion coupled with color and apparent magnitude serve to
separate generally blue, lower luminosity, and larger tangential
velocity WDs from low-metallicity halo subdwarfs (SDs) and main-sequence
stars.

Reduced proper motion is defined as
\begin{equation}
H_m=m+5\log{\mu}+5=M+5\log{V_{\rm{tan}}}-3.379
\label{eqn:rpm}
\end{equation}
where $m$ is apparent magnitude, $M$ is absolute magnitude, $\mu$ is
proper motion in arcseconds yr$^{-1}$, and $V_{\rm{tan}}$ is tangental
velocity in km s$^{-1}$.  In this study, only targets satisfying
$H_g>15.136+2.727(g-i)$ were kept as WD candidates, a reduced proper
motion diagram cut defined by \cite{kilic2006} that eliminates most of
the low-metallicity SDs and virtually all main-sequence stars. A
photometric color cut $J-K_S < 0.5$ was employed to better constrain
initially large samples of WD candidates.  This color cut was adopted
from \cite{subasavage2008} where they found this to be a clear
delimiter between WDs and SD contaminants.  All candidates were
verified to have noticeable proper motions.  These by-eye
verifications were carried out by blinking between digitized POSS I
and POSS II epochs for each target using the Aladin interactive sky
atlas\footnote{\url{http://aladin.u-strasbg.fr/}} \citep{aladin}.

One final model-adherence step was implemented to better probe the
cooler regime (\teff $<$ 7000 K) that is plagued by subdwarf
contaminants.  Remaining targets were crudely compared with WD
synthetic photometry models to estimate \teff values and distances
prior to observations.  Targets poorly represented by the models as
defined below were discarded.  We used pure-hydrogen atmospheric WD
models\footnote{\url{http://www.astro.umontreal.ca/~bergeron/CoolingModels/}}
and assumed log $g=8.0$.  The model grid was cubic spline interpolated
to give a temperature resolution of 10 K.  Both modeled and candidate
photometry were converted to F$_\lambda$ flux values by the
prescription of \cite{holberg}.  The resulting SEDs were normalized by
the $r$ band flux value.  A best fit was determined using a chi-square
minimization between model and target normalized flux values.  If the
photometric error was greater than 0.1 mag then that passband was
ignored for fitting purposes -- this only occurred in the near-IR
passbands as the SDSS magnitudes are significantly better down to our
adopted magnitude limit of $g =$ 19.5.  A $p$ value for
goodness-of-fit was determined using common chi-square lookup tables.
We elected to keep only WD candidates with $p >$ 0.95.

The fitting process provided estimates of \teff and distance for each
target in the final sample.  After spectroscopic observations, all
bona fide WDs were modeled in a more robust fashion for physical
parameter determinations (described in \ref{subsec:model} and
presented here) and should not be confused with this pre-observation
model fitting.  Finally, all previously identified objects were
discarded from our target list, the vast majority of these were
published WDs.

\subsection{Completeness Estimates}
\label{subsec:Complete}

We estimate our target-selection completeness by looking at the
recovered fraction of SDSS-E06 WDs that met our criteria for both
samples (Pass 1 and 2).  For a homogeneous comparison, all of the
relevant DR8 data (i.e., RA, Dec., proper motion, $ugriz$, $JHK_S$)
were extracted for the SDSS-E06 sample by matching plate/fiber/mjd
designations.  A total of 39 objects out of 9316 in SDSS-E06 were not
recovered because the spectra are not included in the DR8 database (hence, 
no plate/fiber/mjd designation).  We confirm that they are available in the DR7 database.  Nevertheless, these few
missing objects should not significantly affect the completion
statistics.

To compare the SDSS-E06 sample (with DR8 data) to our Pass 1 sample,
we implemented identical observational criteria (i.e., proper motion
$>$ 100 mas yr$^{-1}$, $g <$ 19.5, Decl. $< +$30$^\circ$, $J-K_S <$
0.5) to the SDSS-E06 sample.  Our Pass 1 sample was then positionally
cross-matched to this sample and 19 out of 21 were recovered, implying
a completeness of 90\%.  To estimate the completeness of our Pass 2
sample, the SDSS-E06 sample (with DR8 data) was positionally
cross-matched to the LSPM catalog.  For all common objects, identical
observational criteria (i.e., $g < $ 19.5, $J-K_S <$ 0.5) as
implemented in our Pass 2 were applied.  Our Pass 2 sample was
compared via LSPM name to the remaining objects once these criteria
were applied and 31 out of 35 were recovered, implying a completeness
of 89\%.  Note that in both cases, our adopted limiting magnitude
excluded the vast majority of WDs in the SDSS-E06 catalog thereby
leaving us with fairly small numbers with which to estimate
completeness.

To better understand how our selection criteria affects our
completeness, the cuts relating to direct observables (i.e., proper
motion, magnitude, color, and declination) were performed first.
These samples represent the 100\% complete subsample of SDSS-E06
relevant for each of our passes.  We then implement the RPM cut and,
in both cases, recover 100\% of the subsample.  It was only in the
model-adherence cut that recovery rates fell below 100\%.  Because
only a total of 5 unique WDs were not recovered (one unrecovered
object belonged to both samples), we looked more carefully at each one
to ensure our model-adherence criteria were robust.  WD 0756$+$437 is
a magnetic DA with a field strength of $\sim$300 MG
\citep{2009A&A...506.1341K}.  WD 1156$+$132 is classified as DQpec
with deep swan-band carbon absorption.  WD1311$+$129 is classified as
a DBA and has the largest hydrogen abundance of the
\cite{2011ApJ...737...28B} sample of DBAs [log (H/He) = -2.90].  The
remaining two WDs, WD 1559$+$534 and SDSS J101436.01+422622.0 appear
to be normal DA white dwarfs that only marginally failed our
model-adherence criteria with $p > $ 0.9.  It stands to reason that the
first three would fail our model-adherence criteria because their SEDs
are not well-represented by a pure-H model atmosphere.  Thus, while
our completeness suffered because of our model-adherence criteria, the
survey was better able to successfully probe the parameter spaces
dominated by contaminants (\teff $<$ 7000 K) to identify nearby WDs as
discussed in \ref{subsec:nearby}.

\section{Data and Observations}

\subsection{Astrometry and Nomenclature}
\label{astrometry}

For new discoveries, we determine WD names in the conventional manner
\citep[fully described in][]{subasavage2007} by using the target's
epoch 1950 equinox 1950 coordinates.

As stated previously, proper motions were initially extracted from the
SDSS DR7 database and consisted of combined USNO-B1.0 plus SDSS
astrometry \citep{munn}.  \cite{munn} demonstrate that the proper
motions derived from combined astrometry improves those contained
within the USNO-B1.0 catalog by $\sim$25\% when compared to bright,
non-moving, spectroscopically-confirmed QSOs.  However,
\cite{kilic2006} found that high proper motion objects with
neighboring sources within 7\arcsec\space were more likely to have
incorrectly measured proper motions.  To remove this source of
contamination \cite{kilic2006} discarded objects with neighbors within
7\arcsec.  We did not implement this criterion to avoid the
possibility of missing true WDs.

 After an initial night of spectroscopic observations at CTIO
 (discussed in Section \ref{subsec:spec}), we realized 14 SD and
 main-sequence contaminates with incorrect proper motion values from
 the SDSS query.  By-eye verification confirmed these to be bona fide
 proper motion objects, but proper motion magnitudes and position
 angles were erroneous, suggesting mismatches when combining USNO-B1.0
 and SDSS astrometry.  To mitigate this effect, we initiated a second
 query using the LSPM-North Catalog \citep{lspm} as a starting point
 for proper motion values and cross-matched with the SDSS DR7 and
 2MASS databases.  LSPM-North objects were not initially verified by
 eye for proper motion confirmation as this was painstakingly done by
 the original authors.  This lead to a significant reduction in
 contaminants.
 
%Thus, we would expect a similar level of erroneous
%proper motions given that we did not discard candidates with
%neighboring sources within 7\arcsec.

Following observations, proper motions for all spectroscopic
confirmations (WDs and contaminants) were double checked using the
SuperCOSMOS Sky Survey \citep[SSS;][]{hambly}.  In cases where the
USNO-B1.0+SDSS and SSS proper motions were discrepant, a by-eye
inspection was performed to confirm position angle. For these objects,
we adopt the proper motions extracted from the SSS.

WD names, alternate names (LSPM-North where available and SDSS
otherwise), epoch 2000 coordinates and adopted proper motions are
listed in Table 1 for both WDs (\emph{top}) and
contaminants (less the WD name, \emph{bottom}).

\subsection{Photometry}
\label{subsec:phot}

The {\em psfMag} values in the optical $ugriz$ passbands were
extracted from the SDSS DR8.  The target selections were performed
using DR7 photometry but in the interim, DR8 was released and thus, we
use these values for SED modeling discussed in Section
\ref{subsec:model}.  We list these values and their corresponding
errors in Table 1.

As a requirement, all candidate WDs had to contain $JHK_S$ data in
2MASS.  Given that the majority of these targets are near the faint
limit of 2MASS, we utilized the UKIRT Infrared Deep Sky Survey
(UKIDSS) DR6 Large Area Survey to supplement $JHK_S$ for new
discoveries where available.  The UKIDSS project is defined in
\cite{2007MNRAS.379.1599L}. UKIDSS uses the UKIRT Wide Field Camera
\citep[WFCAM;][]{2007A&A...467..777C}. The photometric system is
described in \cite{2006MNRAS.367..454H}, and the calibration is
described in \cite{ukidss}. The pipeline processing and science
archive are described in Irwin et al. (2009, in preparation) and
\cite{2008MNRAS.384..637H}.  UKIDSS magnitudes were transformed to the
2MASS system using the prescription of \cite{ukidss}.  The $JHK_S$
values and corresponding errors are listed in Table 1.
Converted UKIDSS photometry is listed whenever errors in $JHK_S$ are
less than 0.05.  Otherwise, the values are from 2MASS.

\subsection{Spectroscopy}
\label{subsec:spec}

Prior to observations, 28 previously unidentified objects had spectra
in the SDSS DR7 database confirming their WD nature.  These objects
are noted in Table 1, but we do not include their
spectra as they are freely available in the SDSS database.

\begin{figure}
\epsscale{1}
\plottwo{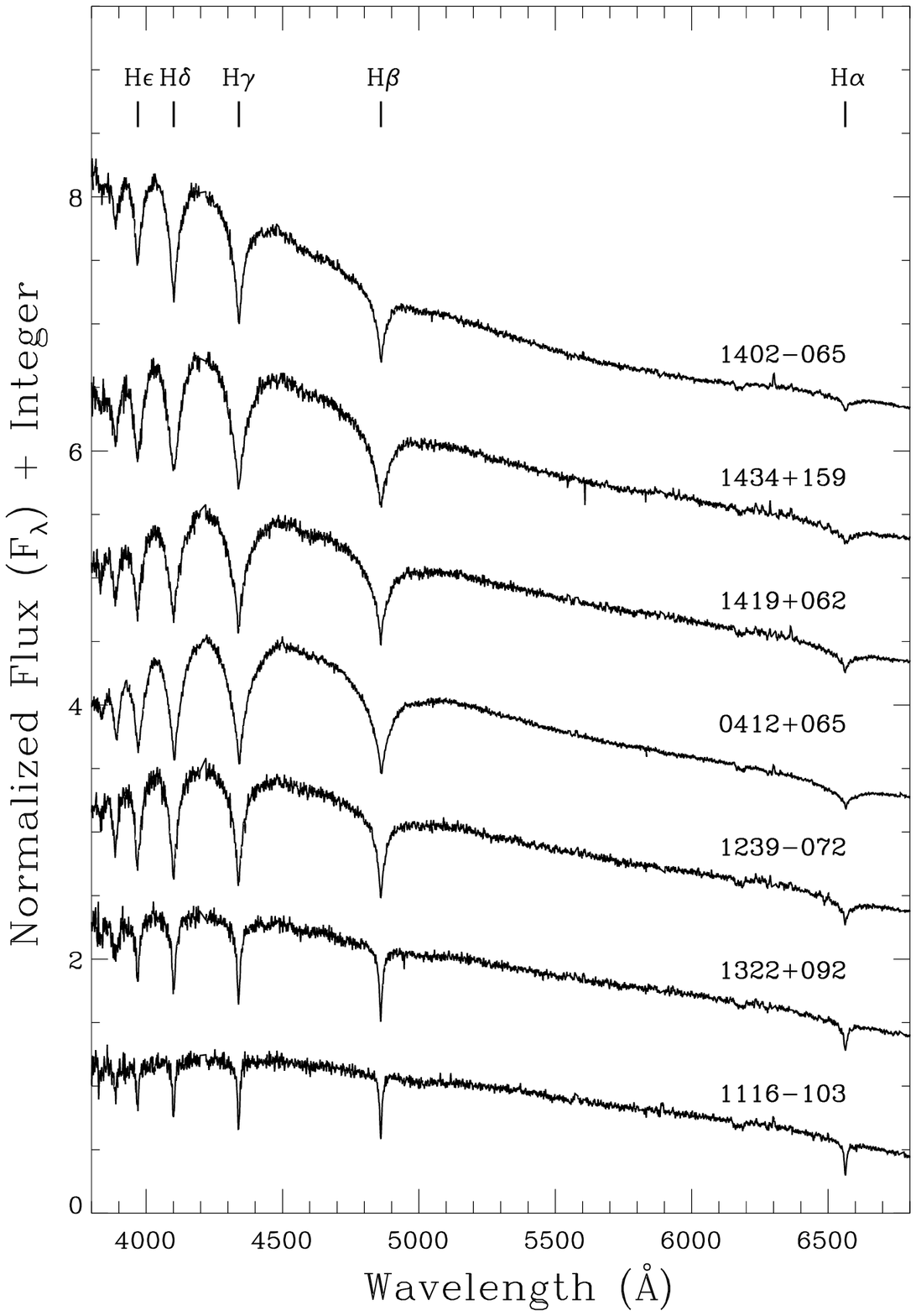}{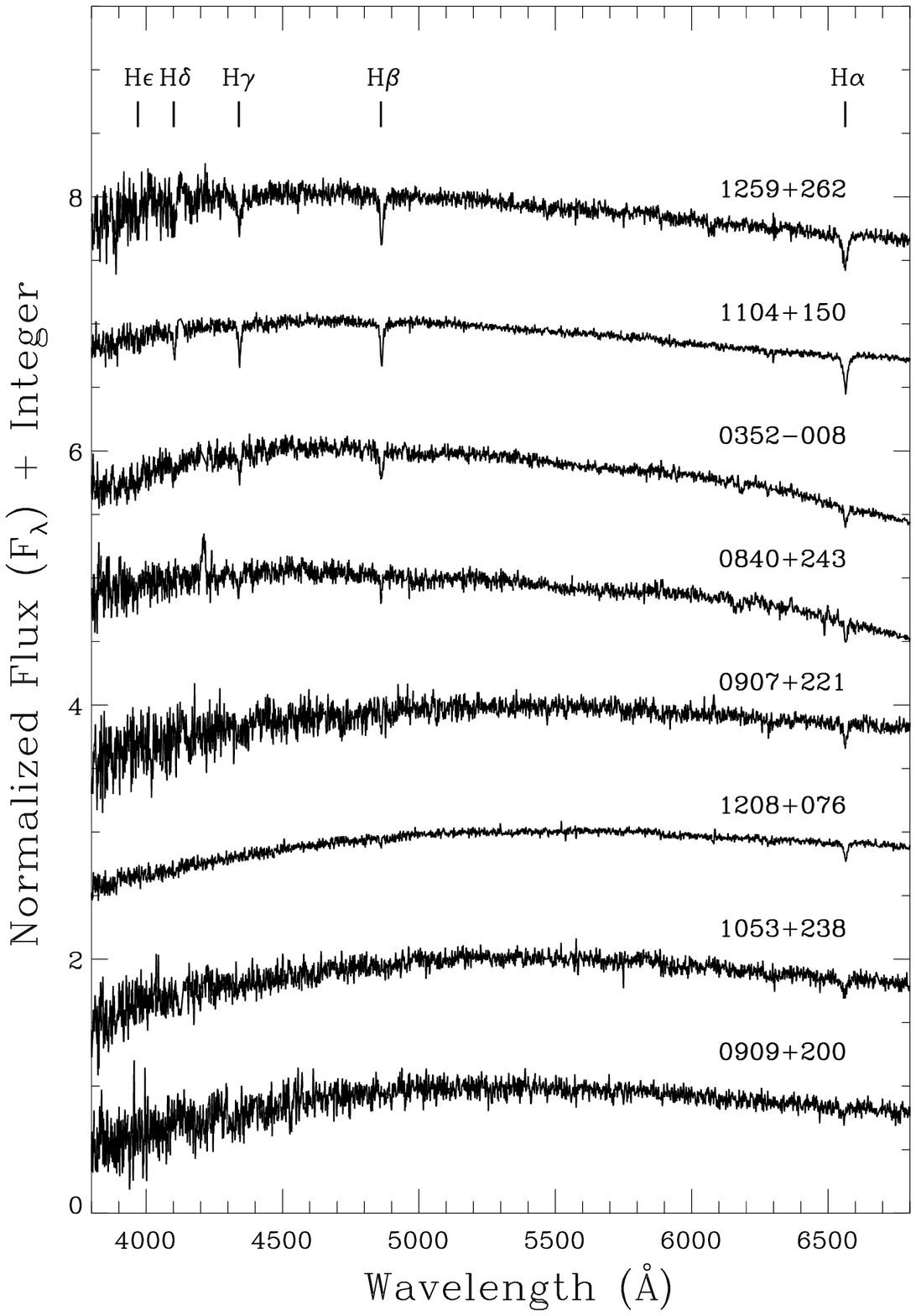}
\caption{Hydrogen-line DA WDs, plotted in order of decreasing \teff
  from top to bottom.  Balmer lines are indicated and WD designations
  are labeled above the spectra.  WD 1419$+$062 is included as
  discussed in Section \ref{subsec:notes}.  Spectra obtained at the CTIO 4m Blanco Telescope.}
\label{fig:4m_da}
\end{figure}

Spectroscopic observations were conducted throughout 2010 and early
2011 from both CTIO and Apache Point Observatory (APO).  At CTIO the
4m Blanco Telescope and Ritchey-Chr{\'e}tien Spectrograph were used.
We selected the KPGL3 grating that covered a wavelength range of
3600-7000 \AA.  Observations in 2010 were taken using a 2$\farcs$0
slit width, oriented due north-south, to provide spectral resolution
of 6 \AA.  Observations in 2011 were taken using a 4$\farcs$0 slit
width to minimize light loss from differential refraction as the slit
was not rotated to the parallactic angle but rather was kept fixed
again due north-south.  With this configuration, the resolution
degraded slightly to 8 \AA.  As can be seen in Figure \ref{fig:4m_da},
which contain spectra from both runs, differential refraction was not
a significant problem during the first run, nor was the loss of
resolution during the second run detrimental for the purpose of
spectral classification.

At APO, the Dual Imaging Spectrograph (DIS) on the ARC 3.5m telescope
was used with the B400+R300 gratings for spectroscopic follow-up.  We
achieved 6 \AA~resolution spanning 3350-9260 \AA, but fringing became
problematic beyond $\sim$7000 \AA.  Red and Blue spectroscopic
channels were reduced independently.  Red channel flux near the
dichroic at approximately 5400 \AA\space was found to be highly
variable, so wavelengths less than 6300 \AA\space were omitted in the
red spectral channel.  The default slit width of 1$\farcs$5 was used
for the DIS spectrograph.

During all of the spectroscopic observing runs, flux standards were
observed each night for flux calibration and HeNeAr lamps were taken
at each telescope pointing for wavelength calibration.  Two spectra of
each target were obtained to permit cosmic ray rejection.  Data were
reduced using standard IRAF\footnote{IRAF is distributed by the
  National Optical Astronomy Observatry, which is operated by the
  Association of Universities for Research in Astronomy, Inc., under
  cooperative agreement with the National Science Foundation}
routines.

\begin{figure}
%\begin{landscape}
%\epsscale{1}
\includegraphics[width=.85\textwidth]{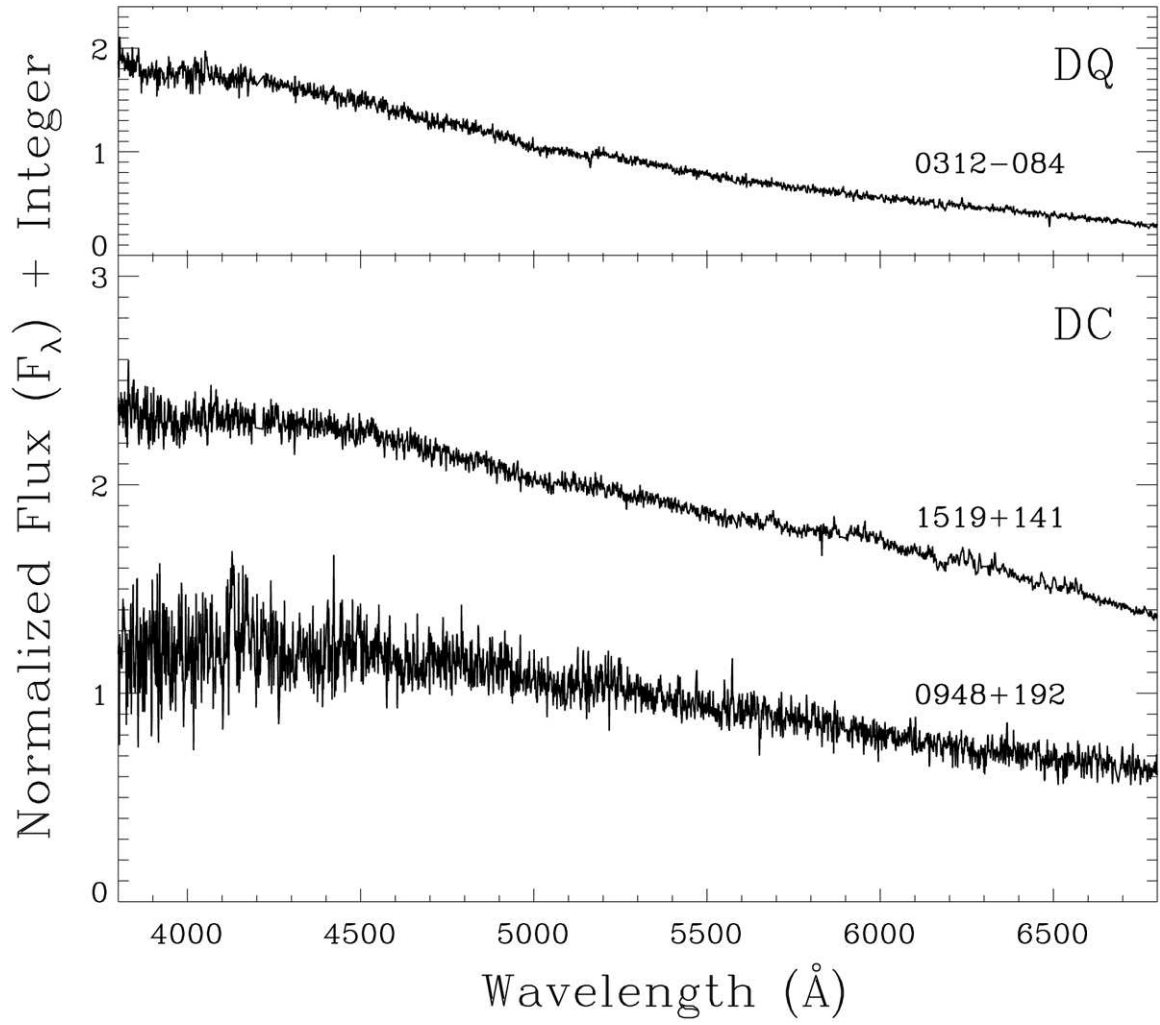}
\caption{DQ ({\it top panel}) and DC ({\it bottom panel})
  WDs, plotted in order of decreasing \teff.  Spectra obtained at the CTIO 4m Blanco Telescope.}
\label{fig:4m_dq_dc}
\end{figure}

\begin{figure}
\epsscale{.75}
\plotone{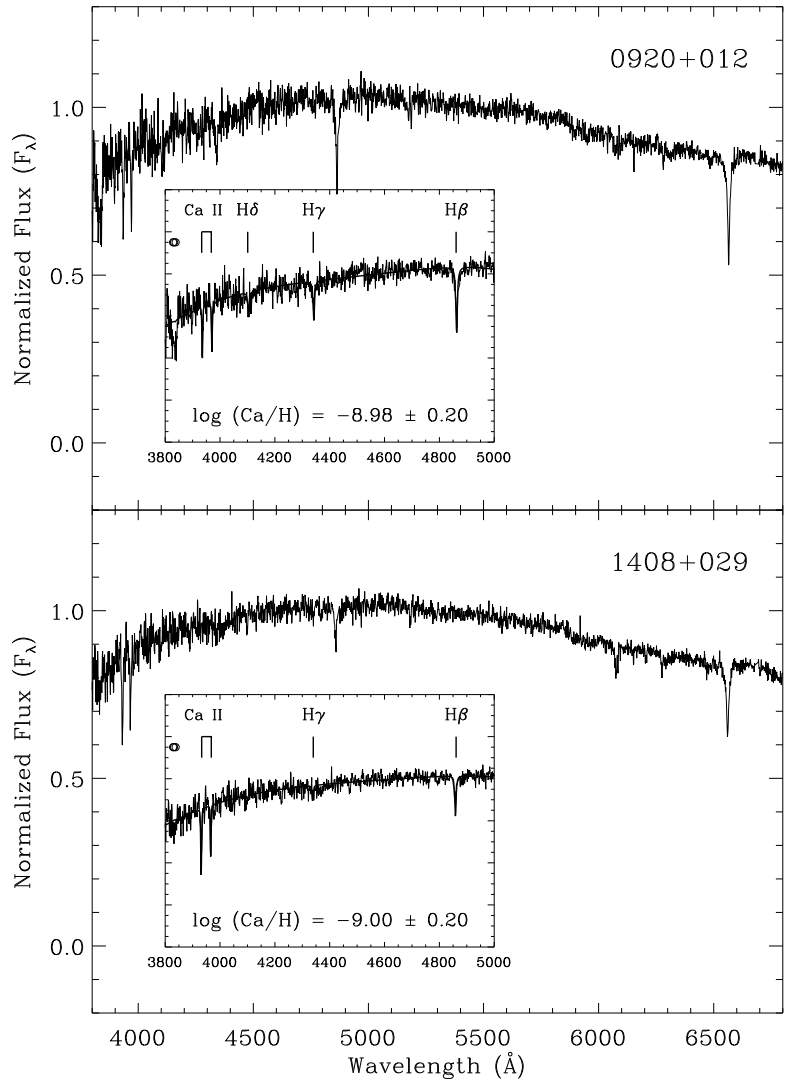}
\caption{DAZ WDs with Ca {\footnotesize II} H \& K features.  Model fits are displayed in the inset panels.  Spectra obtained at the CTIO 4m Blanco Telescope.}\label{fig:4m_daz}
\end{figure}

\begin{figure}
\epsscale{1}
\plottwo{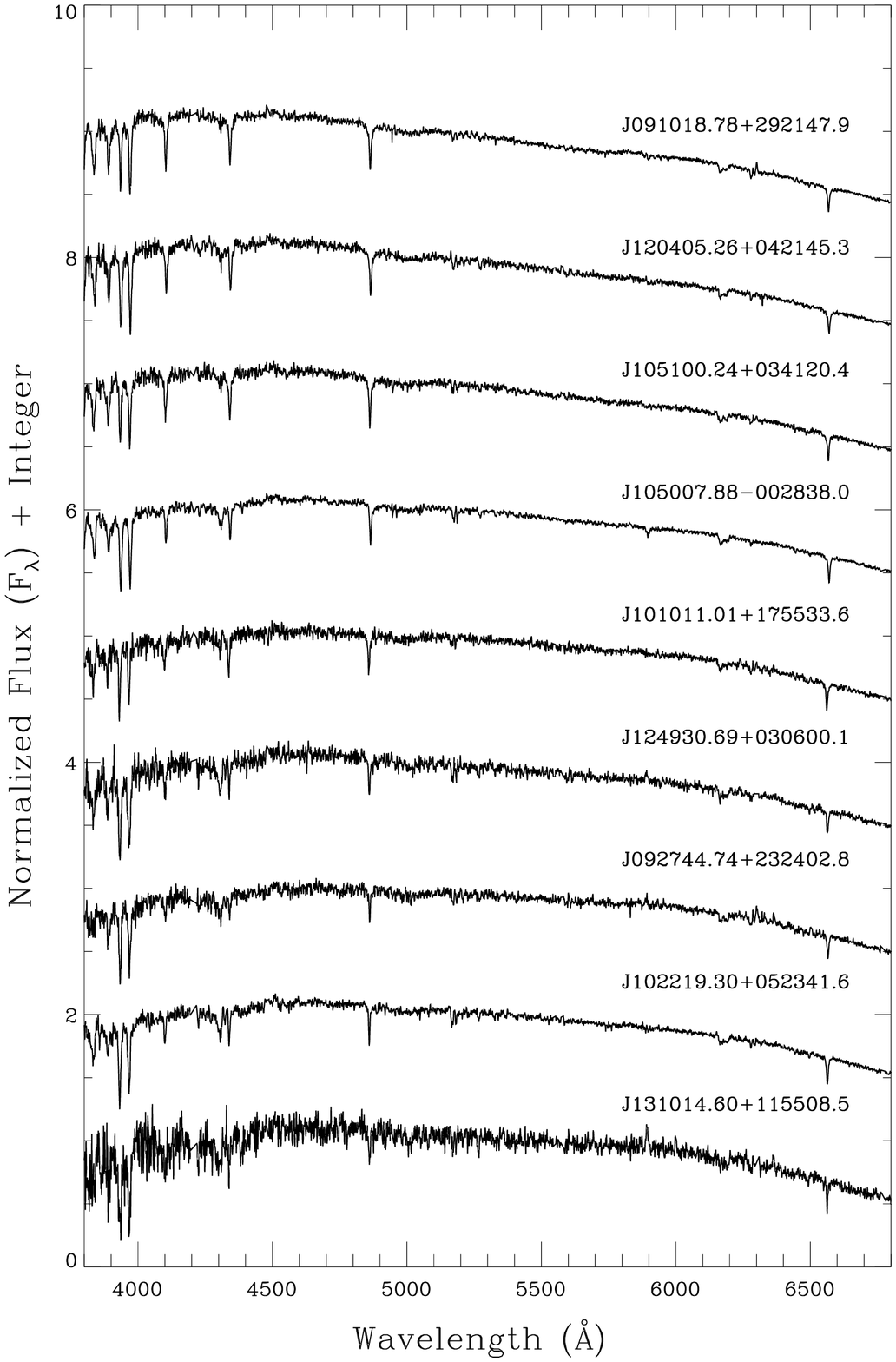}{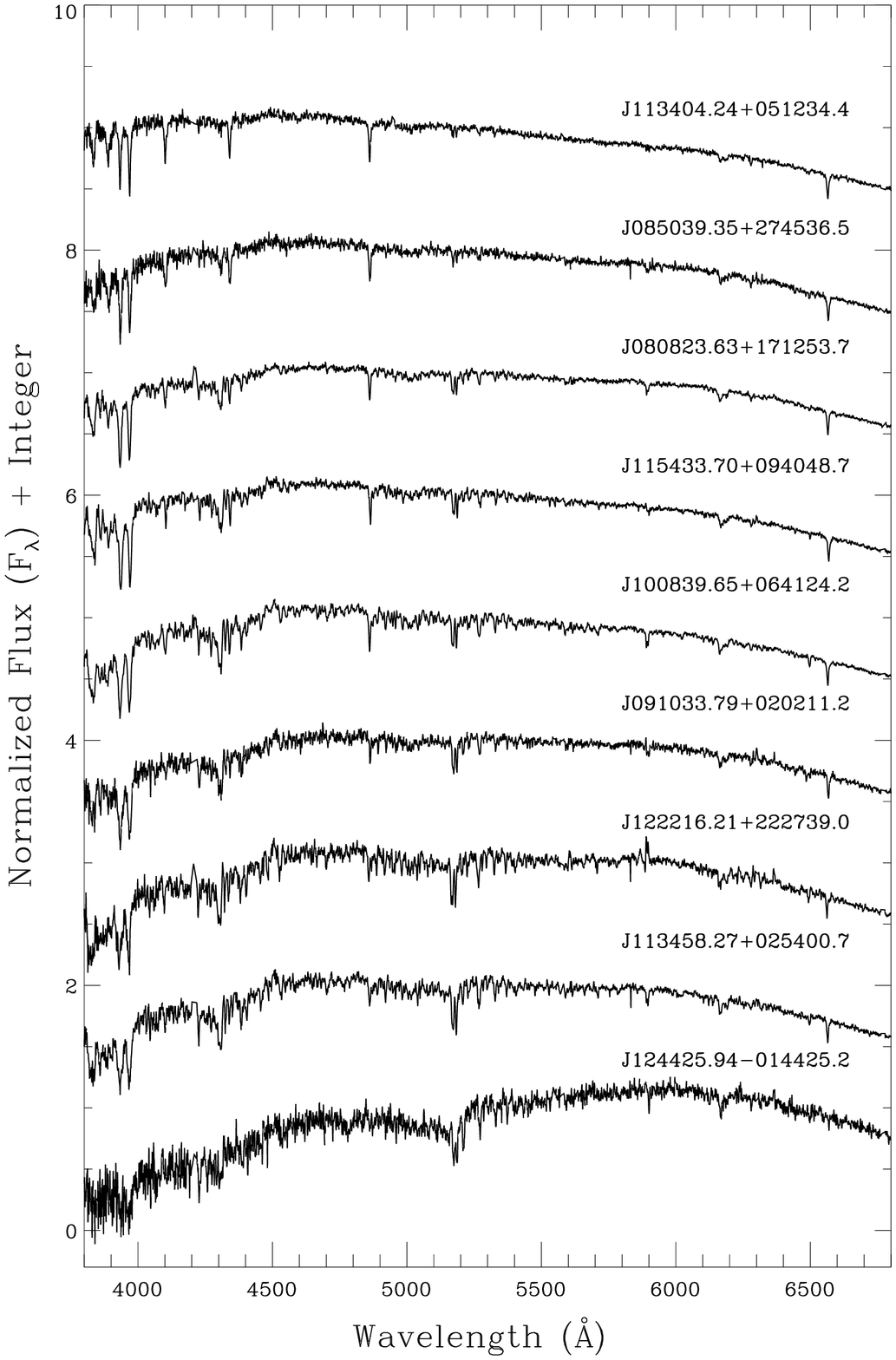}
\caption{Contaminant spectra plotted in
  order of decreasing \teff.  Spectra obtained at the CTIO 4m Blanco Telescope.}
\label{fig:4msd}
\end{figure}

We confirm 19 new WDs from CTIO 4m observations.  Figure
\ref{fig:4m_da} presents 14 DA discoveries and one DA recovery (see
Section \ref{subsec:notes}). Figure \ref{fig:4m_dq_dc} displays one DQ
and two DC WDs.  Figure \ref{fig:4m_daz} shows spectra and model fits
for two DAZ WDs with detectable Ca {\footnotesize II} absorption
features at 3933 and 3968 \AA.  From CTIO, 18 contaminants were
observed and their spectra are plotted in Figure \ref{fig:4msd}.
Defining absorption features in these contaminant spectra are due to
metal and molecular content (Ca {\footnotesize II}, CH, MgH).

\begin{figure}
\includegraphics[angle=90,width=0.95\textwidth]
{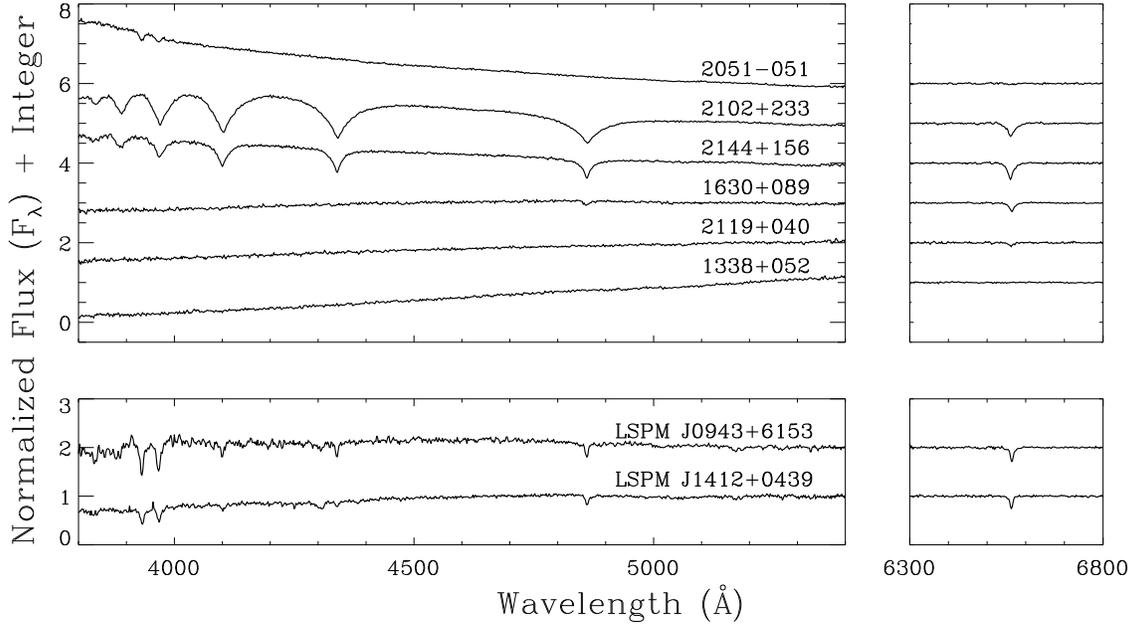}
\caption{Spectra obtained from the ARC 3.5m Telescope at APO.  Left panels represent the blue
  channel and right panels represent a subsection of the red
  channel centered on H$\alpha$.  Top panels are new WD
  discoveries while the bottom panels are two contaminants, likely SDs.
  All spectra are plotted in order of decreasing \teff.  }
\label{fig:3msp}
\end{figure}

We confirm 6 new WDs from APO 3.5m observations: three DA, two DC and
one DZ with Ca {\footnotesize II} absorption in an otherwise
featureless spectrum.  The upper panels of Figure \ref{fig:3msp}, show
these discoveries. We experienced only two contaminants (both likely
SDs) and their spectra are plotted in the lower two panels of Figure
\ref{fig:3msp}.

\section{Analysis}

\subsection{Modeling of Physical Parameters}
\label{subsec:model}

Our model atmospheres for WDs are similar to those described at length
in \citet[][and references therein]{liebert05} and \citet{BSW95}, with
several improvements discussed in \cite{TB07} and \cite{TB11}. In
particular, we now make use of the improved Stark broadening profiles
for the hydrogen lines developed by \citet{TB09}.  Our models for DQ
and DZ stars, which include metals and molecules in the equation of
state and opacity calculations, are described in detail in
\citet{duf05} and \citet{duf07}, respectively.

Table 2 contains SED-derived \teff (column 2) and distance
(column 4) for WDs with pure H or pure He atmospheres (denoted in
column 3).  A complete discussion of our SED-fitting procedure can be
found in \cite{2001ApJS..133..413B}.  For "polluted" WDs (DAZ, DQ,
DZ), these parameters rely on the iterative, combined SED and spectral
fitting procedures defined in \cite{duf05,duf07}.  Metal abundances
derived from the model fits for these targets are listed in the notes
section of Table 2.  For all cases, we assume log $g$ = 8.0.
Spectral subtypes (column 5) are determined for the DA WDs using the
temperature index of \cite{mccook}, where the temperature index equals
50,400/\teff.  In addition, spectroscopic line profile fitting was
performed as described in \cite{bsl} for all DAs (and one DBA -- WD
1457$+$249) with sufficient line absorption and signal-to-noise to
produce a reliable fit.  These results are listed in columns 6 and 7
of Table 2.

\subsection{Metal-Polluted DA White Dwarfs}

Figure \ref{fig:4m_daz} displays spectra and model fits for WD
0920$+$012 and WD 1408$+$029. These two DA WDs exhibit Ca
{\footnotesize II} H \& K features and earn the classification of DAZ.
Hypotheses to explain these spectral features include enrichment from
(1) the interstellar medium or, (2) debris disk accretion, with the
latter being heavily favored \citep{2010ApJ...714.1386F}.

%In the cases of DAZs, the settling
%timescales for metals in cool ($T_{\rm{eff}} <$ 20,000 K) hydrogen-rich
%WD atmospheres is on the order of a few years \citep{2009A&A...498..517K},
%strongly suggesting that these objects are accreting at the current
%epoch.  Thus, DAZ WDs are best explained by steady accretion from a
%debris disk that is likely the remnant of a tidally disrupted (and
%subsequently destroyed) rocky body.  

WD 0920$+$012 is estimated to be at 33.4 pc, while WD 1408$+$029 is
estimated to be at 26.5 pc, and both are modeled to have a log
(Ca)/(H) = -9.0.  Given their relatively bright apparent magnitudes
($g \sim$ 17), these targets would be excellent candidates for
follow-up IR studies to possibly detect emission from the accretion
disk and better characterize the system.

\subsection{ZZ Ceti White Dwarfs}
\label{subsec:zzceti}

\begin{figure}
\includegraphics[angle=90,width=0.95\textwidth]
{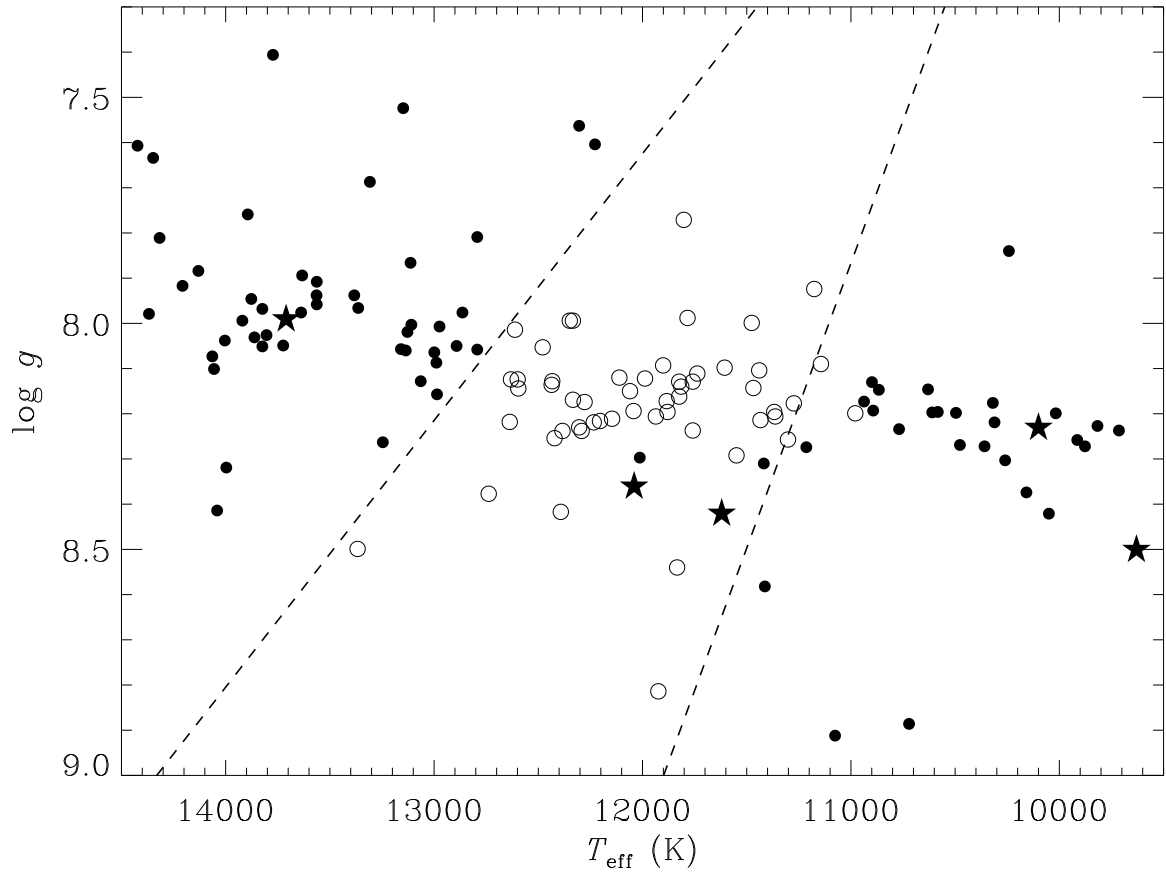}
\caption{Plot of select new DA WDs with spectroscopic determinations of log $g$ and \teff
  ({\it filled stars}) - from left to right: 0412+065, 2102+233, 1419+062, 1239+072, 1257+185.  Also plotted are WDs that are not observed to
  vary ({\it filled circles}) and ZZ Ceti confirmations ({\it
    open circles}).  The dashed lines represent the instability strip
  boundaries as defined by \cite{zzstrip2}.}
\label{fig:zzceti}
\end{figure}

Two of our confirmed WDs were modeled to lie in the ZZ Ceti
instability strip: WD 1419$+$062 and WD 2102$+$233.  Figure
\ref{fig:zzceti} shows the instability strip as recently redefined by
\cite{zzstrip2} using improved model spectra with the new Stark
profiles described above.

Differential photometry was performed on both ZZ Ceti candidates.
Candidate flux was normalized by bright non-varying stars in the same
field to obtain a differential light curve.  The light curve was then
normalized by the mean value to obtain fractional variation about the
mean.  Frequency content was analyzed using the magnitude of the
Fourier Transform (FT) of the time series data.

\begin{figure}
\epsscale{1}
\plotone{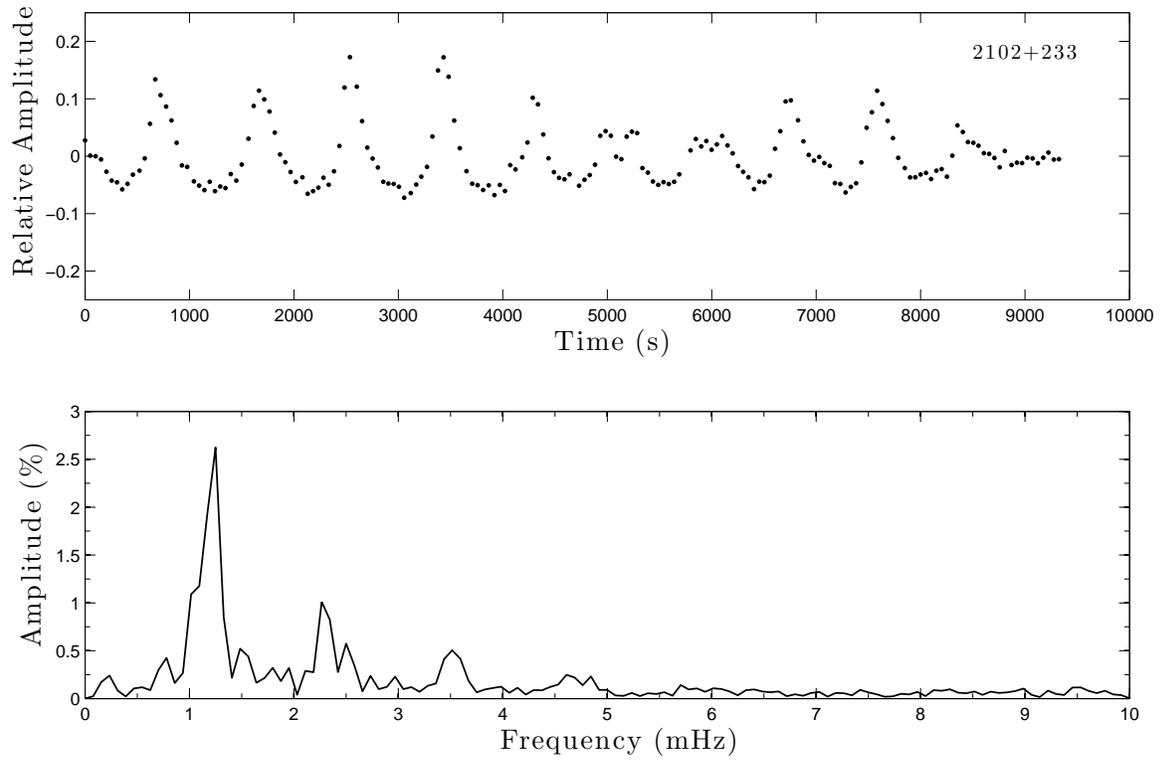}
\caption{Light curve (top panel) and FT (bottom panel) for WD2102+233.  Object was observed from the CTIO 0.9m telescope at a cadence of $\sim$ 50 seconds.  From the FT, the dominant mode was found to be $\sim$ 800 seconds}\label{fig:zz1}
\end{figure}

WD 2102$+$233 was observed using the CTIO 0.9m telescope using the
full 13\farcm6 field and the BG 40 filter.  The time cadence was
$\sim$50 seconds.  As can be seen in Figure \ref{fig:zz1}, we identify
a dominant pulsational period at $\sim$800 seconds with an amplitude of
$\sim$2.6\%.

\begin{figure}
\epsscale{1}
\plotone{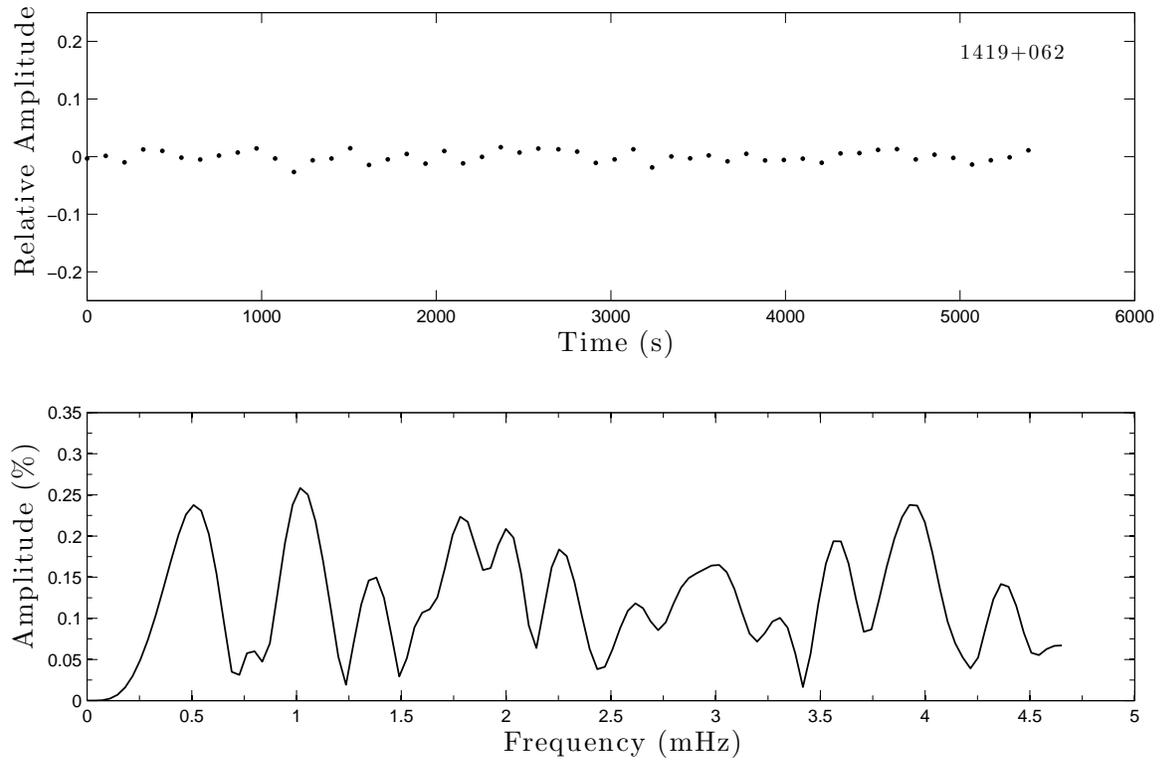}
\caption{Light curve (top panel) and FT (bottom panel) for WD1419+062.  Object was observed from the CTIO 1.0m telescope at a cadence of $\sim$ 100 seconds.  From the FT, no dominant mode was observed above the noise level out to the Nyquist frequency.}\label{fig:zz2}
\end{figure}

WD 1419$+$062 was observed using the CTIO 1.0m telescope, with a
20\farcm0 field, and the BG 40 filter. We detect no obvious pulsations
from these observations (see Figure \ref{fig:zz2}).  From the FT, the
observed noise level is $\sim$0.25\%, with no obvious peaks.  With our
less-than-optimal sampling rate ($\sim$100 seconds), we could only
hope to detect periods $>$ 200 seconds based on the Nyquist sampling
theorem, so this object may be pulsating outside our detection
sensitivity.  However, considering its location in Figure
\ref{fig:zzceti} lying near the red edge of the instability strip, we
would expect a longer period variable.

\subsection{Nearby White Dwarfs}
\label{subsec:nearby}

We find three new WDs with distance estimates within the 25 pc horizon
of interest (WD 1338$+$052 at 13.7 $\pm$ 2.7 pc, WD 1630$+$089 at 13.2
$\pm$ 2.3 pc, and WD 2119$+$040 at 22.1 $\pm$ 3.6 pc) that has been
adopted from the Catalog of Nearby Stars \citep[CNS;][]{cns} and the
NStars Database \citep{henry03}.  \cite{holberg_nearby} determine that
the 20 pc WD sample is $\sim$80\% complete based on the assumption
that the 13 pc WD sample is complete.  Two of these WDs, and possibly
the third if it is slightly less distant than expected, will lie
within 20 pc if proximity is confirmed.  Moreover, WD 1338$+$052 and
WD 1630$+$089 are modeled to lie at a distance very near to 13 pc.  If
either object proves to lie within 13 pc, the local WD population must
be denser than previously thought.  With any new WD discoveries within
13 pc, a constant-density extrapolation would increase the amount of
missing WDs in the 13 pc to 20 pc range.  The exact current
completeness statistics will depend on more robust distance
determinations.

These three objects are being observed for trigonometric parallaxes
via the Cerro Tololo Inter-American Observatory Parallax Investigation
\citep[CTIOPI;][]{jao,henry,subasavage,riedel,2011AJ....141..117J}
program to confirm proximity.  To the best of our knowledge, the two
WDs estimated to be within 20pc are the nearest WDs discovered using
SDSS data, if trigonometric parallaxes confirm proximity.  Continuing
discoveries of the coolest WDs, especially those nearby and thus
suitable for parallax measurements, will provide anchor points for WD
atmospheric models that are vital to widely-used cosmic dating
techniques.
 
\subsection{Notes in Individual Systems}
\label{subsec:notes}

{\bf WD 0351$-$002} is also known as SA 95-42 and is a popular
spectrophotometric standard star.  \cite{1990AJ.....99.1621O} lists
this object along with 24 other spectrophotometric standard
candidates, including spectra.  For reasons unclear, this is one of
three objects that do not have a spectral type in Table 1 of
\cite{1990AJ.....99.1621O} yet the spectrum published in that same
work shows broad Balmer absorption indicative of a DA WD.  We include
it here as a new discovery.

{\bf WD 0412$+$065} is also known as GD 59.  It was classified as a WD
suspect by \cite{1965LowOB...6..155G}, but we found no spectroscopic
confirmation in the literature even though it is listed as a WD in
Simbad.  We include it here as a new discovery.

{\bf WD 1402$+$065} is also known as PG 1402$+$065 and was previously
labeled as a subdwarf in the catalogs of \cite{1986ApJS...61..305G}
and subsequently included in the subdwarf catalog of
\cite{1988SAAOC..12....1K}, based on the previous determination.  Our
spectrum (see Figure \ref{fig:4m_da}) clearly shows broad Balmer
absorption and thus, we include it here as a new WD discovery.  In
fact, this WD is the hottest of those spectroscopically observed from
CTIO plotted in Figure \ref{fig:4m_da} ($T_{\rm eff}$ = 26,190 K).

{\bf WD 1419$+$062} is also known as PG 1419$+$062 and was first
published by \cite{1980ApJ...238..685G} as a DA.  However, it appears
in \cite{1986ApJS...61..305G} with a "sd" designation indicative of
being a hot subdwarf.  It then appears in the catalog of
\cite{1988SAAOC..12....1K}, again classified as a hot subdwarf based
on the \cite{1986ApJS...61..305G} designation.  Here we confirm it to
be a bona fide DA WD and suspect the \cite{1986ApJS...61..305G}
designation is a typo.  We include the spectroscopy and the modeled
parameters yet do not classify this object as a new discovery.  This
object is also discussed in Section \ref{subsec:zzceti} as a ZZ Ceti
candidate.

{\bf WD 1434$+$159} is also known as GD 168.  It was classified as a
WD suspect by \cite{1965LowOB...6..155G}, but we found no
spectroscopic confirmation in the literature even though it is listed
as a WD in Simbad.  We include it here as a new discovery.

{\bf WD 1457$+$249} has a spectrum dominated by He and was SED-modeled
using a pure He model atmosphere.  During the spectroscopic line
fitting analysis, it became evident that, because of weak Balmer
H$\beta$ absorption, trace amounts of H existed in the atmosphere.
Thus, this object is classified as a DBA with the best-fitting model
including log (H/He) = -5.8 $\pm$ 0.1.

\section{Discussion}
\label{sec:disc}

We present and characterize 54 WDs.  Of these, 26 objects were
spectroscopically observed from either CTIO or APO (including
recovered known object WD 1419$+$062, as discussed in Section
\ref{subsec:notes}), while the remaining 28 objects have spectra
available in the SDSS DR7 database for a total of 53 new WDs.  It is
likely the SDSS DR7 spectra will be contained in the forthcoming DR7
White Dwarf Catalog described by \cite{kleinmandr7}.  Additionally, 21 objects overlap (17 from SDSS spectra and 4 photometrically-selected) between this study and a recent publication by \cite{girven} to identify DA WDs in SDSS.

We find three WDs expected to be in the local neighborhood, with the
closest (WD1630$+$089) expected to be 13.2 pc distant.  Twenty-one WDs
are modeled to have \teff $<7000$ K, with four of those objects
modeled to have \teff $<5000$ K.  We also confirm WD 2102$+$233 as a
new ZZ Ceti star.

\begin{figure}
\epsscale{1}
\plotone{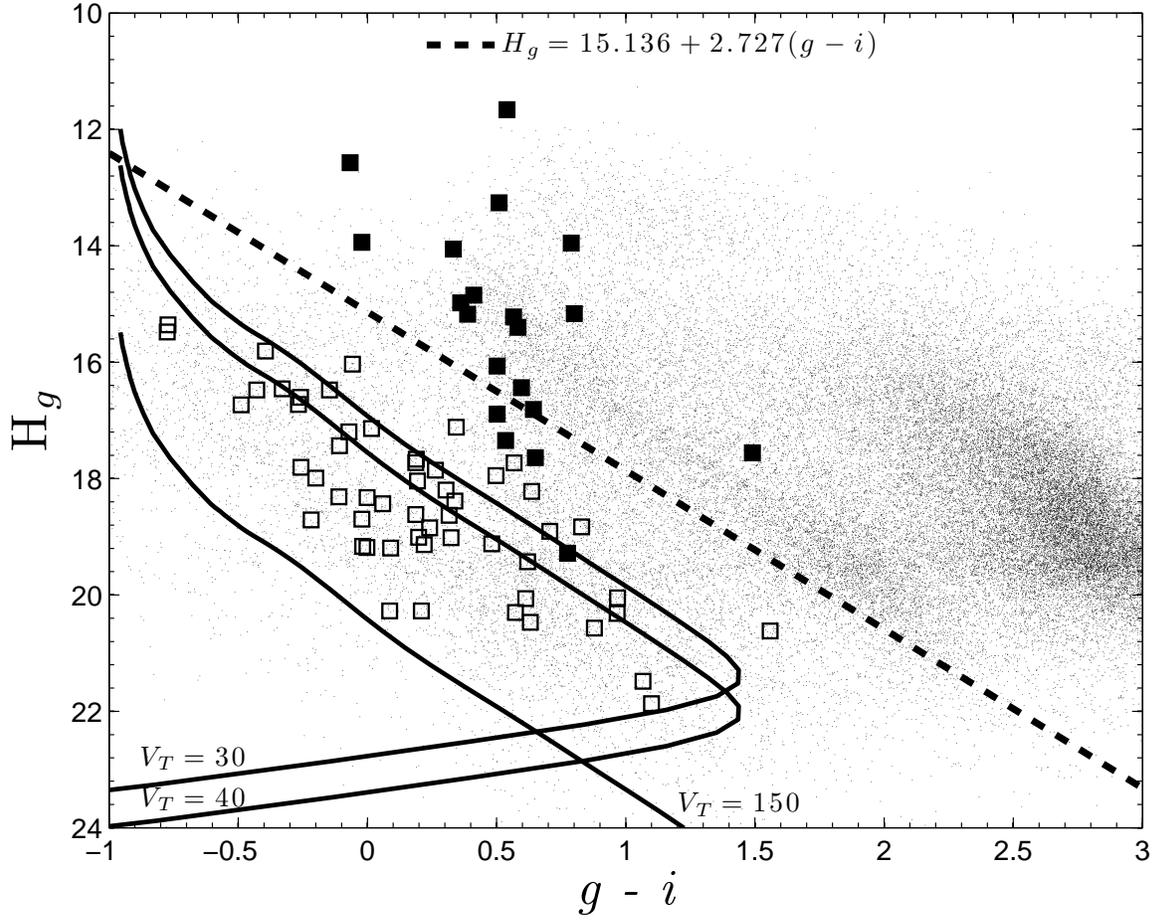}
\caption{Reduced proper motion diagram showing WDs from this study ({\it
    open squares}) and contaminants ({\it filled squares}).  The
  dashed line represents the cut adopted from \cite{kilic2006} above
  which targets were discarded.  Contaminants above the curve are
  discussed in Section \ref{sec:disc}.  Modeled cooling curves for
  pure-hydrogen WDs with log $g$ = 8 and $V_{\rm{tan}} =$ 30, 40, and
  150 km s$^{-1}$ are plotted as solid curves.  Background stars were taken from SDSS for reference.}
\label{fig:rpm}
\end{figure}

Our survey is summarized in Table 3.  Briefly, our
selection criteria were designed to optimally identify cooler WDs in
parameter spaces where contaminants dominate.  As such, we realized
some contamination in our final sample, both identified by our spectra
and also from the literature (Table 3 column 10)
as listed by Simbad.  Our methodology was initially impacted by
problematic proper motions leading to significant contamination by
subdwarfs and main-sequence stars (Pass 1 in Table
3).  Once correct proper motions are applied to these
objects (see Figure \ref{fig:rpm}), it is clear the contamination
would have been largely avoided as these objects fall above our RPM
cut.  Our corrected contamination is generally consistent with
\cite{kilic2010} who show decreasing contamination rates with
increasing $V_{\rm{tan}}$ cutoffs. They quote a 1.3\% contamination
rate for $V_{\rm{tan}} \geq 30$ km s$^{-1}$ and indeed, we find one
contaminant scattered within the $V_{\rm{tan}} \geq 30$ km s$^{-1}$
cutoff.  While we show yet another example of the effectiveness of RPM
to isolate WDs, there is an inherent bias against the slowest moving
WDs.  It is unlikely that this bias will be completely removed until a
magnitude-limited astrometric survey is conducted, such as $Gaia$ or
$LSST$.

Only a subset of our WD candidates were able to be observed because of
telescope time constraints.  The last column in Table
3 shows the number of targets left unobserved.
Furthermore, all of our targets were constrained to the SDSS footprint
thereby covering only a fraction of the sky (Table 3
column 3) that is largely weighted towards northern declinations.  For
these reasons, we expect that more WDs of interest remain
undiscovered.  Highly anticipated photometric surveys such as
Pan-STARRS and the Dark Energy Survey (DES) will certainly aid in the
discovery of additional WDs.  With this in mind, we publish our
observed contaminates to add to those already known to have similar
photometric properties and proper motions to WDs
\citep[e.g.,][]{kilic2006,kilic2010}.  By identifying large samples of contaminants now, disentangling them from WD candidates may be
easier in the future using empirical and statistical methods.
\cite{2011MNRAS.tmp..545N} provide an example of statistical
techniques based on photometry to classify supernovae.  For similar
techniques to work in our context, large robust training sets of both
SD and WD exemplars are needed before any statistical assertions can
be made, and applied to large data sets.

\acknowledgments

We are grateful to the anonymous referee, who provided detailed
comments and suggestions that enabled the manuscript to be more
concise and clear.  C.S.~and J.P.S.~wish to thank the 2010 CTIO REU
Program, of which C.S.~was a student (the majority of this work was
conducted during the program under the direction of J.P.S.).  In
particular, we thank Chris Smith and Nicole van der Bliek for
director's discretionary time on the 4m Blanco Telescope at CTIO and
Suzanne Hawley for director's discretionary time on the APO ARC 3.5m
telescope.  We thank Bart Dunlap for spectral analysis tips, and Adam
Kowalski for DIS reduction advice.

This work was, in part, based on observations obtained with the Apache
Point Observatory 3.5-meter telescope, which is owned and operated by
the Astrophysical Research Consortium.  This work is supported in part
by the NSERC Canada and by the Fund FQRNT (Qu\'ebec). PD is a CRAQ
postdoctoral fellow.

This publication makes use of
data products from the Two Micron All Sky Survey, which is a joint
project of the University of Massachusetts and the Infrared Processing
and Analysis Center/California Institute of Technology, funded by the
National Aeronautics and Space Administration and the National Science
Foundation.

Funding for the SDSS and SDSS-II has been provided by the Alfred
P. Sloan Foundation, the Participating Institutions, the National
Science Foundation, the U.S. Department of Energy, the National
Aeronautics and Space Administration, the Japanese Monbukagakusho, the
Max Planck Society, and the Higher Education Funding Council for
England. The SDSS Web Site is \url{http://www.sdss.org/}.  The SDSS is
managed by the Astrophysical Research Consortium for the Participating
Institutions. The Participating Institutions are the American Museum
of Natural History, Astrophysical Institute Potsdam, University of
Basel, University of Cambridge, Case Western Reserve University,
University of Chicago, Drexel University, Fermilab, the Institute for
Advanced Study, the Japan Participation Group, Johns Hopkins
University, the Joint Institute for Nuclear Astrophysics, the Kavli
Institute for Particle Astrophysics and Cosmology, the Korean
Scientist Group, the Chinese Academy of Sciences (LAMOST), Los Alamos
National Laboratory, the Max-Planck-Institute for Astronomy (MPIA),
the Max-Planck-Institute for Astrophysics (MPA), New Mexico State
University, Ohio State University, University of Pittsburgh,
University of Portsmouth, Princeton University, the United States
Naval Observatory, and the University of Washington.

\clearpage
%\begin{landscape}
%table 1 placeholder:
\begin{deluxetable}{c}
\tabletypesize{\tiny}
\startdata
\enddata
\end{deluxetable}

\begin{figure}
\centering
\hspace{-1.5in}
%\epsscale{1.5}
\includegraphics[ scale=.6, angle=90]{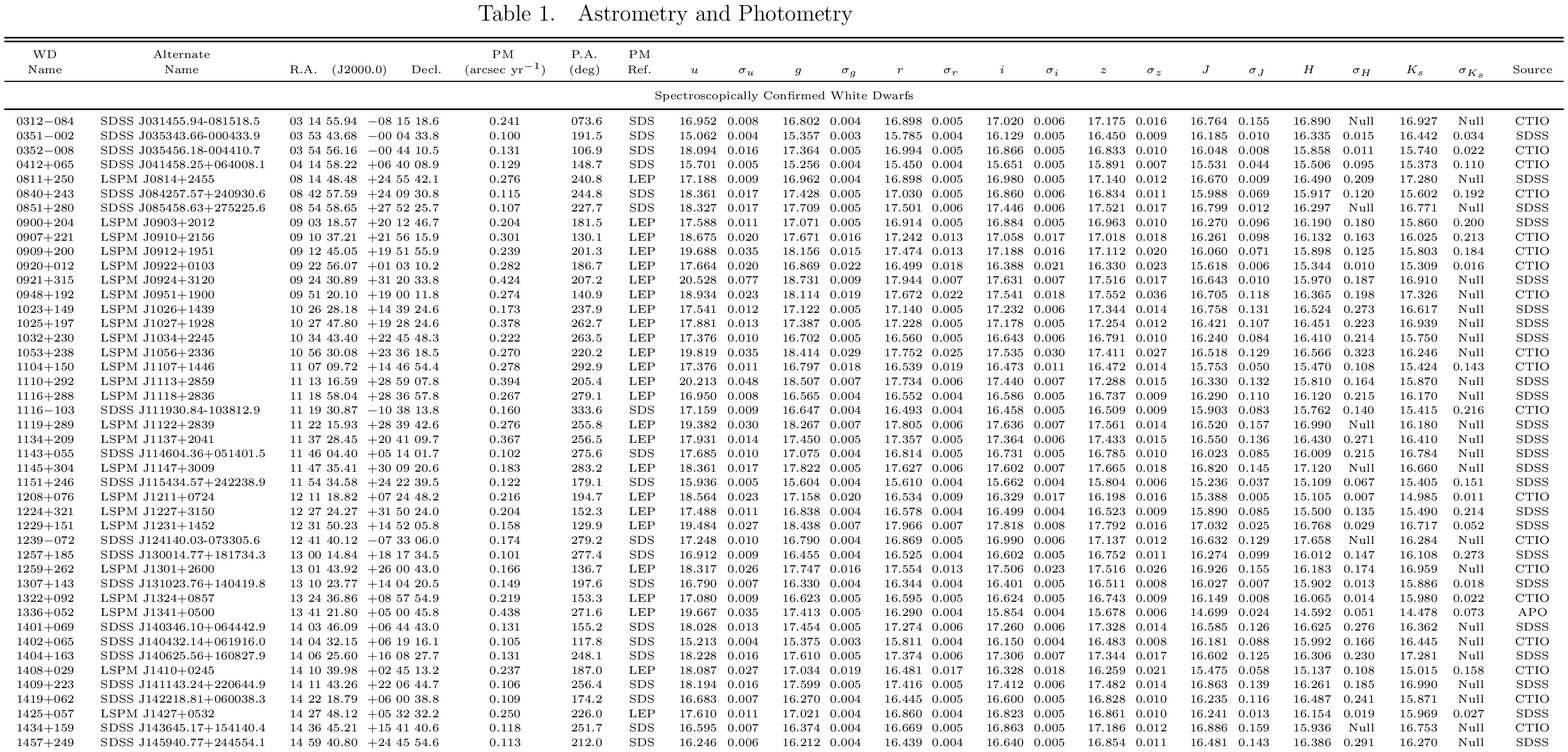}
\end{figure}
\clearpage
%\end{landscape}

\clearpage
%\begin{landscape}
\begin{figure}
\centering
%\epsscale{1.5}
\hspace{-1.5in}
\includegraphics[scale=.6, angle=90]{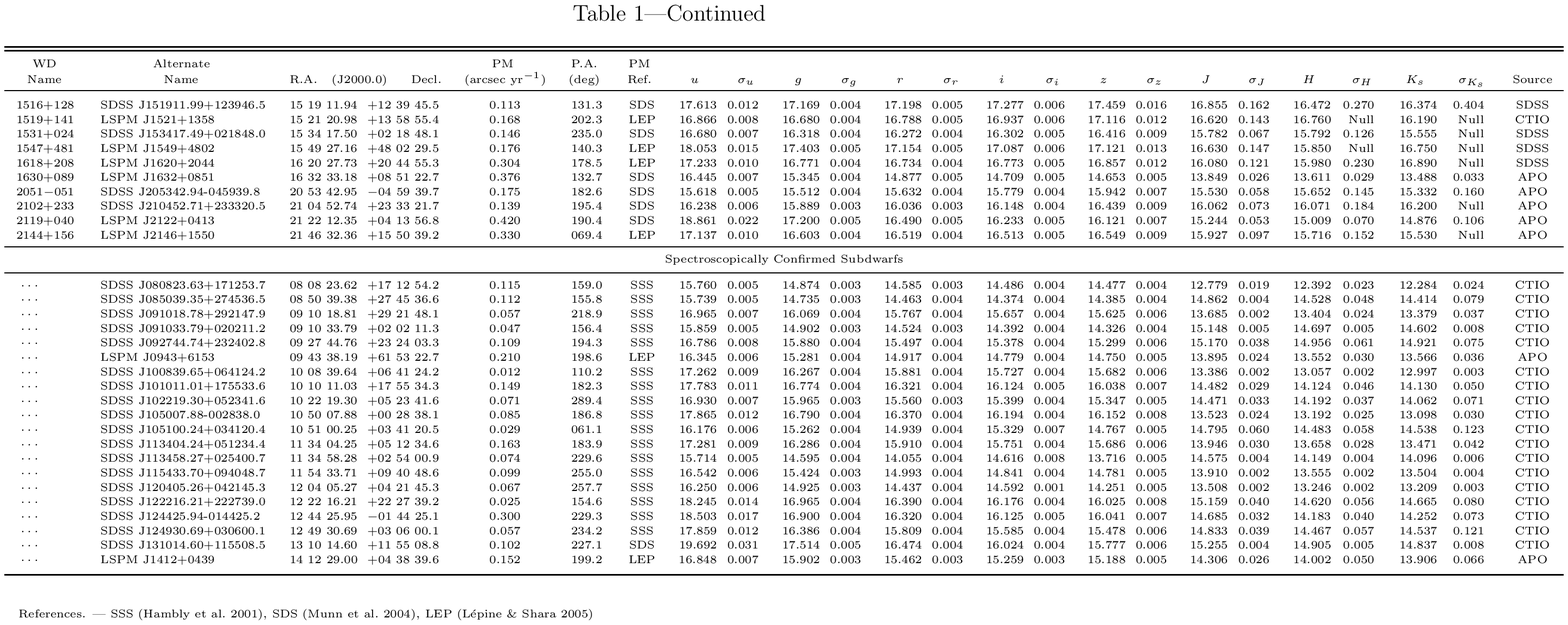}
\end{figure}
\clearpage
%\end{landscape}
\clearpage

\hoffset-30pt{}
\begin{deluxetable*}{p{60pt}r@{ $\pm$ }lcr@{ $\pm$ }lcl@{ $\pm$ }rr@{ $\pm$ }lc}
\tabletypesize{\tiny} 
\tablecaption{Derived Parameters for New White Dwarfs
\vspace{-10pt}
\label{char}}
\tablewidth{0pt} \tablehead{\colhead{WD}        &
                            \multicolumn{2}{c}{Photometric}&
                            \colhead{}          &
                            \multicolumn{2}{c}{Dist.}&
                            \colhead{Spec.}     &
                            \multicolumn{4}{c}{Spectroscopic} &
                            \colhead{}         \\
			    \colhead{Name}       &
                            \multicolumn{2}{c}{$T_{\rm{eff}}$ (K)}&
                            \colhead{Comp.}      &
                            \multicolumn{2}{c}{(pc)}&
                            \colhead{Type}      &
                            \multicolumn{2}{c}{$T_{\rm{eff}}$ (K)} &
                            \multicolumn{2}{c}{log $g$} &
                            \colhead{Notes}\\
                            \colhead{(1)}    &
                            \multicolumn{2}{c}{(2)}   &
                            \colhead{(3)}    &
                            \multicolumn{2}{c}{(4)}   &
                            \colhead{(5)}    &
                            \multicolumn{2}{c}{(6)}   &
                            \multicolumn{2}{c}{(7)}   &
                           \colhead{(8)}\\[-3pt] }

\startdata
\\[-5pt]
0312$-$084\dotfill  &  9080  &   50  &  He(+C)   &   68.9  &  11.2  &   DQ  & \multicolumn{4}{c}{\nodata}                 & \tablenotemark{c}    \\
0351$-$002\dotfill  &  29,480  &  260  &   H   &  129.5  &  26.3  &   DA1.5  & 36,960  & 190 & 7.63 & 0.03                           &  \tablenotemark{a}   \\
0352$-$008\dotfill  &   6160  &   30  &   H   &   40.1  &   6.7  &   DA7.0  & \multicolumn{4}{c}{\nodata}      &   \tablenotemark{b}  \\
0412$+$065\dotfill  &  11,920  &  180  &   H   &   53.6  &   9.5  &   DA4.0  & 13,720  & 280 &  7.99 & 0.03                            &     \\
0811$+$250\dotfill  &   8220  &   40  &  He(+C)   &   60.9  &  9.9  &   DQ  &   \multicolumn{4}{c}{\nodata}                &   \tablenotemark{d}  \\
0840$+$243\dotfill  &   5950  &   30  &   H   &   38.0  &   6.2  &   DA7.5  & \multicolumn{4}{c}{\nodata}    &   \tablenotemark{b}  \\
0851$+$280\dotfill  &   7030  &   40  &   H   &   63.5  &  10.7  &   DA6.0  & 6780 & 90 & 7.93 & 0.17                              &     \\
0900$+$204\dotfill  &   7360  &   40  &   H   &   52.2  &   8.9  &   DA6.0  & 7190 & 40 & 7.92 & 0.06                               &     \\
0907$+$221\dotfill  &   5830  &   80  &   H   &   40.3  &   6.6  &   DA7.5  &  \multicolumn{4}{c}{\nodata}              &   \tablenotemark{b}  \\
0909$+$200\dotfill  &   4950  &   70  &   H   &   32.0  &   5.3  &   DA9.0  & \multicolumn{4}{c}{\nodata}                 &   \tablenotemark{b}  \\
0920$+$012\dotfill  &   6290  &   60  & H(+Ca)&   33.4  &   5.6  &   DAZ & \multicolumn{4}{c}{\nodata}                  &  \tablenotemark{e}   \\
0921$+$315\dotfill  &   4810  &   60  &  He   &   38.1  &   7.3  &   DC  &   \multicolumn{4}{c}{\nodata}                &     \\
0948$+$192\dotfill  &   5850  &  100  &  He   &   51.5  &   8.5  &   DC  & \multicolumn{4}{c}{\nodata}                  &     \\
1023$+$149\dotfill  &   9190  &   90  &   H   &   83.1  &  14.2  &   DA5.5  & 9180 & 50 & 8.06 & 0.06                              &     \\
1025$+$197\dotfill  &   7330  &   40  &   H   &   59.5  &  10.2  &   DA6.0  &  7140 & 70 & 8.34 & 0.10                             &     \\
1032$+$230\dotfill  &   6770  &   20  &He(+Ca)&   42.2  &   6.8  &   DZ  &   \multicolumn{4}{c}{\nodata}                &   \tablenotemark{f}  \\
1053$+$238\dotfill  &   5110  &  110  &   H   &   39.4  &   6.4  &   DA9.0  & \multicolumn{4}{c}{\nodata}                 &   \tablenotemark{b}  \\
1104$+$150\dotfill  &   6720  &  110  &   H   &   37.8  &   6.2  &   DA6.5  & 6710 & 90 & 7.94 & 0.20                              &    \\
1110$+$292\dotfill  &   4810  &   80  &  He   &   34.7  &   7.1  &   DC  &   \multicolumn{4}{c}{\nodata}                &     \\
1116$+$288\dotfill  &   8850  &   70  &   H   &   59.4  &  10.2  &   DA5.5  & 8550 &  30 & 8.37 & 0.04                              &     \\
1116$-$103\dotfill  &   7350  &   40  &   H   &   42.8  &   7.3  &   DA6.0  & 7360 & 50 &  7.77 & 0.09                               &     \\
1119$+$289\dotfill  &   5690  &   40  &   H   &   50.1  &   8.2  &   DA8.0  & \multicolumn{4}{c}{\nodata}    &   \tablenotemark{b}  \\
1134$+$209\dotfill  &   7860  &   60  &   H   &   71.1  &  12.3  &   DA5.5  & 7740 & 40 & 8.01 & 0.06                              &     \\
1143$+$055\dotfill  &   6740  &   40  &   H   &   43.1  &   7.0  &   DA6.5  & 6840 & 60 & 7.91 & 0.10                              &     \\
1145$+$304\dotfill  &   7060  &   40  &  He   &   69.4  &  11.5  &   DC  &  \multicolumn{4}{c}{\nodata}                 &     \\
1151$+$246\dotfill  &   9060  &   60  &   H   &   39.7  &   6.6  &   DA5.5  & 8700 & 30 & 8.74 & 0.03                               &      \\
1208$+$076\dotfill  &   5430  &   40  &   H   &   25.5  &   4.4  &   DA8.5  & \multicolumn{4}{c}{\nodata}               &    \tablenotemark{b}  \\
1224$+$321\dotfill  &   6730  &   40  &   H   &   38.7  &   6.3  &   DA6.5  & 6530 & 80 & 7.99 & 0.15                              &     \\
1229$+$151\dotfill  &   5840  &   40  &   H   &   57.3  &   9.4  &   DA7.5  & \multicolumn{4}{c}{\nodata}    &    \tablenotemark{b} \\
1239$-$072\dotfill  &   9970  &  100  &   H   &   82.9  &  13.7  &   DA5.0  & 10,100 & 50 & 8.23 & 0.05                              &      \\
1257$+$185\dotfill  &   9500  &   70  &   H   &   65.4  &  11.1  &   DA5.5  & 9630 & 30 & 8.50 & 0.03                              &      \\
1259$+$262\dotfill  &   7050  &  150  &   H   &   65.1  &  10.9  &   DA6.0  & \multicolumn{4}{c}{\nodata}                 &    \tablenotemark{b}   \\
1307$+$143\dotfill  &   9000  &   40  &   H   &   55.8  &   9.6  &   DA5.5  & 8610 & 30 & 8.21 & 0.04                              &       \\
1322$+$092\dotfill  &   8440  &   40  &   H   &   56.3  &   9.7  &   DA5.5  & 8210 & 40 & 8.11 & 0.07                              &       \\
1338$+$052\dotfill  &   4360  &   50  &  He   &   13.7  &   2.7  &   DC  &    \multicolumn{4}{c}{\nodata}               &       \\
1401$+$069\dotfill  &   7140  &   50  &  He   &   60.0  &   9.9  &   DC  &   \multicolumn{4}{c}{\nodata}                &         \\
1402$+$065\dotfill  &  26,190  &  210  &   H   &  116.6  &  23.1  &   DA2.0  & 27,490 & 170 & 7.83 & 0.03                             &   \tablenotemark{a}     \\
1404$+$163\dotfill  &   6810  &   50  &   H   &   56.9  &   9.2  &   DA6.5  & 6560 & 100 & 7.65 & 0.02                              &        \\
1408$+$029\dotfill  &   5570  &   90  & H(+Ca)&   26.5  &   4.3  &   DAZ & \multicolumn{4}{c}{\nodata}                  &   \tablenotemark{g}     \\
1409$+$223\dotfill  &   7160  &   50  &  He   &   64.5  &  10.7  &   DC  &    \multicolumn{4}{c}{\nodata}               &        \\
1419$+$062\dotfill  &  11,210  &  100  &   H   &   78.8  &  13.6  &   DA4.5  & 11,620 & 110 & 8.42 & 0.04                            &         \\
1425$+$057\dotfill  &   7060  &   30  &  He   &   48.5  &   8.1  &   DC  &   \multicolumn{4}{c}{\nodata}                &         \\
1434$+$159\dotfill  &  15,690  &  170  &   H   &  112.8  &  20.9  &   DA3.0  & 17,940 & 170 & 8.06 & 0.03                             &         \\
1457$+$249\dotfill  &  12,820  &   90  &  He   &   85.6  &  14.8  &   DBA  & 13,510 & 100 & 8.09 & 0.08                              &   \tablenotemark{h}      \\
1516$+$128\dotfill  &   9290  &   80  &   H   &   86.6  &  14.6  &   DA5.5  & 9290 & 40 & 8.12 & 0.05                              &         \\
1519$+$141\dotfill  &  10,160  &   60  &  He   &   78.6  &  13.0  &   DC  &  \multicolumn{4}{c}{\nodata}                &         \\
1531$+$024\dotfill  &   8510  &   60  &   H   &   49.1  &   8.5  &   DA5.5  & 8220 & 30 & 8.61 & 0.04                              &          \\
1547$+$481\dotfill  &   6780  &   40  &  He   &   51.9  &   8.6  &   DC  &  \multicolumn{4}{c}{\nodata}                 &          \\
1618$+$208\dotfill  &   8490  &   60  &   H   &   61.3  &  10.7  &   DA5.5  & 9070 & 30 & 8.16 & 0.04                              &         \\
1630$+$089\dotfill  &   5740  &   20  &   H   &   13.2  &   2.2  &   DA8.0  & \multicolumn{4}{c}{\nodata}      &       \tablenotemark{b}  \\
2051$-$051\dotfill  &  10,470  &   60  &He(+Ca)&   48.1  &   8.0  &   DZ  &    \multicolumn{4}{c}{\nodata}              &  \tablenotemark{i}       \\
2102$+$233\dotfill  &  11,080  &  110  &   H   &   64.5  &  10.5  &   DA4.5  & 12,040 & 0.08 & 8.36 & 0.02                           &          \\
2119$+$040\dotfill  &   5150  &   50  &   H   &   22.1  &   3.6  &   DA9.0  & \multicolumn{4}{c}{\nodata}      &       \tablenotemark{b}   \\
2144$+$156\dotfill  &   7730  &   50  &   H   &   47.0  &   8.1  &   DA5.5  & 8340 & 40 & 8.49 & 0.05                              &          \\

\enddata
\tablenotetext{a}{$T_{\rm eff}$ from the SED fit is unreliable for the hottest WDs.}
\tablenotetext{b}{A combination of weak Balmer lines and/or noisy spectra prohibited a reliable spectral fit.}
\tablenotetext{c}{Best fit model includes log (C/He) = -4.2 $\pm$ 0.2.}
\tablenotetext{d}{Best fit model includes log (C/He) = -5.1 $\pm$ 0.2.}
\tablenotetext{e}{Best fit model includes log (Ca/H) = -9.0 $\pm$ 0.2.}
\tablenotetext{f}{Best fit model includes log (Ca/He) = -9.5 $\pm$ 0.2.}
\tablenotetext{g}{Best fit model includes log (Ca/H) = -9.0 $\pm$ 0.2.}
\tablenotetext{h}{Best fit model includes log (H/He) = -5.9 $\pm$ 0.1}
\tablenotetext{i}{Best fit model includes log (Ca/He) = -10.5 $\pm$ 0.2.}

\end{deluxetable*}

\clearpage

\begin{deluxetable*}{ccccccccccc}

\tablecaption{Survey Summary \label{summarytable}
}
\tabletypesize{\tiny} 

\tablehead{\colhead{Samp.} & 
  \colhead{PM} & 
  \colhead{Sky\tablenotemark{c}} & 
  \colhead{All Objs.} & 
  \colhead{Selected\tablenotemark{d}} &
  \colhead{\# Obs.} &
  \colhead{\# Obs.} &
  \colhead{\# SDSS} &
  \colhead{\# Publ.} &
  \colhead{\# Publ.} &
  \colhead{Remaining} \\
  \colhead{} & 
  \colhead{Source} &  
  \colhead{\%} & 
  \colhead{N} & 
  \colhead{N} & 
  \colhead{WDs}  & 
  \colhead{contam.} & 
  \colhead{Spect.} & 
  \colhead{WDs} & 
  \colhead{OTHER}  & 
  \colhead{N} \\
  \colhead{(1)}  &
  \colhead{(2)}  &
  \colhead{(3)}  &
  \colhead{(4)}  &
  \colhead{(5)}  &
  \colhead{(6)}  &
  \colhead{(7)}  &
  \colhead{(8)}  &
  \colhead{(9)}  &
  \colhead{(10)} &
  \colhead{(11)} 
}

\startdata
Pass 1\tablenotemark{a} &  USNO & 23 & 97137 & 1028  & 19  & 18 & 12 & 105 & 15 & 859 \\

Pass 2\tablenotemark{b} &  LSPM & 29 & 19408 & 190  & 6  & 2 & 16 & 71 & 9 & 86 \\
\enddata
\tablenotetext{a}{Initial Constraints: PM $>$ 0.10 arcsec yr$^{-1}$, Decl. $< +30$ deg., $g <$ 19.5, 2MASS match, SDSS DR8 footprint}
\tablenotetext{b}{Initial Constraints: PM $>$ 0.15 arcsec yr$^{-1}$, Decl. $> 0$ deg., $g <$ 19.5, 2MASS match, SDSS DR8 footprint}

\tablenotetext{c}{Estimated by applying initial declination constraints to SDSS DR8 footprint.}
\tablenotetext{d}{Target selection criteria (see section \ref{sec:TargSel}): RPM cut, $J-K_s$ cut, model-adherence cut.}

\end{deluxetable*}

%%%%%%%%Tables end

\clearpage

\clearpage
%\end{landscape}

\clearpage

\clearpage

\clearpage

\clearpage

\clearpage

\clearpage

\end{document}